\documentclass[aip,rsi,reprint]{revtex4-1}
\usepackage[pdftex]{graphicx} 
\usepackage{amsmath,amssymb} %
\usepackage{pifont}
\renewcommand{\pi}[0]{\textrm{\Pisymbol{psy}{"70}}} 
\newcommand{\rmu}[0]{\textrm{\Pisymbol{psy}{"6D}}} 

\begin{document}
\title{Phase noise measurement by zero-crossing analysis with a double recorder setup in the radiofrequency range}
\author{Makoto Takeuchi}
\email{takeuchi@phys.c.u-tokyo.ac.jp}
\affiliation{Graduate School of Arts and Sciences, The University of Tokyo, Tokyo, 153-8902, JAPAN}
\author{Haruo Saito}
\affiliation{Graduate School of Arts and Sciences, The University of Tokyo, Tokyo, 153-8902, JAPAN}

\begin{abstract}
The phase noise of low-noise oscillators is conventionally measured by the cross-spectrum method (CSM), which has a complicated setup. We developed an alternative method called zero-crossing analysis with a double recorder setup (ZCA-DRS) that has much simpler configuration, which we previously demonstrated to measure phase noise in the audible frequency range. In this study we conducted experiments using ZCA-DRS to measure phase noise in the radiofrequency range. A temperature-compensated crystal oscillator was used to generate a periodic signal at 27 MHz. The results demonstrated that  the obtained single-side band spectrum was almost the same as that obtained by CSM. The measurement sensitivity was limited by the jitter of the internal clock and noise of the analog-to-digital converter. Thus, ZCA-DRS can be used as an alternative to CSM for phase noise measurement in the radiofrequency range. 
\end{abstract}


\maketitle 

\section{Introduction}
The phase noise of low-noise oscillators is conventionally measured by the cross-spectrum method (CSM), which has a complicated setup. Fig. 1 (a) shows a conceptual diagram of the CSM.\cite{Rubiola10} 
\begin{figure}[ht]
\includegraphics{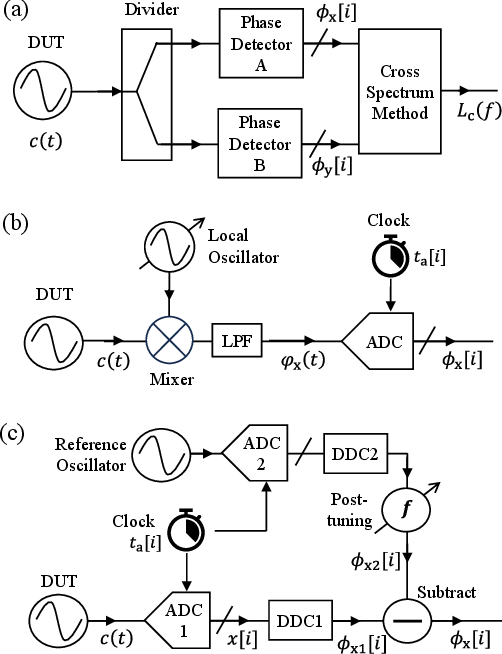}
\caption{\label{Fig:conventional}Conventional phase-noise measurement: (a) Setup of the cross-spectrum method (CSM). The output signal of the device under test (DUT) is measured by two independent phase detectors, and the phase noise is estimated from the correlated components. (b) Phase detector based on the frequency down-conversion (FDC) circuit comprising a local oscillator (LO), mixer, low-pass filter (LPF), clock, and analog-to-digital converter (ADC). (c) Phase detector based on a software-defined radio (SDR) circuit comprising a reference oscillator, clock, and two ADCs.}
\end{figure}
The device under test (DUT) outputs a continuous sinusoidal voltage signal $c(t)$:
\begin{align} 
c(t)=A_0\left\{1+\alpha(t)\right\}\cos\left\{\omega t+\theta_0+\varphi_\mathrm{c}(t)\right\}, \label{Eq:c} 
\end{align}
where $A_0$, $\omega$, and $\theta_0$ are the amplitude, angular frequency, and initial phase, respectively, of the pure sinusoidal signal. The parameters $\alpha(t)$ and $\varphi_\mathrm{c}(t)$ represent amplitude modulation (AM) and phase modulation, respectively. The output signal $c(t)$ is divided into two lines by a divider, which are fed into two independent phase detectors. The phase detectors digitally sample the phase $\varphi_\mathrm{c}(t)$ and record it as $\phi_\mathrm{x}[i]$ and $\phi_\mathrm{y}[i]$. If the noise of the phase detectors are negligible and their sampling timings are simultaneous, then $\phi_\mathrm{x}[i]=\phi_\mathrm{y}[i]=\varphi_\mathrm{c}(t_\mathrm{a}[i])$, where $i$ is an integer and $t_\mathrm{a}[i]=iF_\mathrm{s}^{-1}$. In practice, $\phi_\mathrm{x}[i]\neq\phi_\mathrm{y}[i]$ because the phase detectors are not ideal, and their sampling timings are not simultaneous.  CSM involves calculating the components $\phi_\mathrm{x}[i]$ and $\phi_\mathrm{y}[i]$ that vary in the common mode to estimate $\varphi_\mathrm{c}(t)$ and obtaining the single sideband (SSB) spectrum of the DUT phase noise $L_\mathrm{c}(f)$. When the number of measurements is small or the power spectral density is very low, careful estimation is necessary.\cite{Baudiquez2020,Gruson2020} 

As shown in Fig. \ref{Fig:conventional}(a), CSM requires two independent phase detectors. Fig. \ref{Fig:conventional}(b) depicts the internal structure of the phase detector more concretely. For simplicity, the local oscillator (LO) is pretuned to output a sine wave with the same frequency and initial phase as $c(t)$. The output of the LO is adjusted by a phase lock loop, which is an important technique that determines the performance of a phase detector.\cite{Papez2013,SALZENSTEIN2016118} The output signals of the DUT and LO are multiplied by a mixer. The mixer outputs the angular frequency $2\omega$, for which components are blocked by a low-pass filter (LPF) to result in and the transmitted signal $\varphi_\mathrm{x}(t)$. The circuit between $c(t)$ and $\varphi_\mathrm{x}(t)$ is known as a frequency down-conversion (FDC) circuit. If the FDC circuit operates ideally, then $\varphi_\mathrm{x}(t)=\varphi_\mathrm{c}(t)$. Finally, the analog signal $\varphi_\mathrm{x}(t)$ is digitally sampled to obtain $\phi_\mathrm{x}[i]$. The sampling timing of the analog-to-digital converter (ADC) is controlled by the clock. If there is no sampling jitter or ADC noise, then $\phi_\mathrm{x}[i]=\varphi_\mathrm{x}(t_\mathrm{a}[i])$. In practice, the output of a standalone phase detector contains FDC circuit noise,\cite{Homayoun2013} sampling jitter, and ADC noise.\cite{Olaya2017} These noises can be eliminated by using another independent phase detector and CSM. 

Fig. 1(c) illustrates the internal structure of a direct digital phase detector that has been recently developed.\cite{Grove2004,Imaike2017} The output signal of the DUT is digitally sampled by ADC1, which is driven by the clock. The recorded waveform $x[i]$ is converted to phase data $\phi_\mathrm{x1}[i]$ via digital down-conversion (DDC). The circuit between $c(t)$ and $\phi_\mathrm{x1}[i]$ is known as software-defined radio (SDR).\cite{Sherman2016} If there is no sampling jitter or ADC noise, then $\varphi_\mathrm{c}(t_\mathrm{a}[i])=\phi_\mathrm{x1}[i]$. The sampling jitter in $\phi_\mathrm{x1}[i]$ is observed by driving another SDR using the same clock (ADC2 and DDC2). If the angular frequency of the reference oscillator is different from $\omega$, post-tuning is performed to convert the output of DDC2 to phase data $\phi_\mathrm{x2}[i]$ to the sampling jitter at angular frequency $\omega$. Subtracting $\phi_\mathrm{x1}[i]$ from $\phi_\mathrm{x2}[i]$ yields the output of the direct digital phase detector $\phi_\mathrm{x}[i]$. The direct digital phase detector improves upon the conventional phase detector by removing sampling jitter. However, the noises of ADC1 and ADC2 cannot be removed, so CSM is still required to measure the phase noise of the DUT. 

We have developed a method for measuring the phase noise of a digital audio player called zero-crossing analysis with a double recorder setup (ZCA-DRS).\cite{Takeuchi2023} Phase noise can be observed in the time domain, which eliminates the need for the design, development, and calibration of low-noise FDC circuits.  ZCA-DRS has a very simple structure, and it does not require a FDC circuit or reference oscillator. It has fewer circuit components than a phase-noise analyzer based on two direct digital phase detectors and thus should have lower fabrication costs. In our previous study, we were able to demonstrate that ZCA-DRS could measure phase noise in the audible frequency range by using two digital audio recorders with an analog/digital (A/D) resolution of 24 bit and sampling rate of 192 kSa/s.\cite{Takeuchi2023} Following this success, we expected that ZCA-DRS could be extended to the radiofrequency (RF) range by replacing the digital audio recorders with digital oscilloscopes. Here, we report the phase noise measurement of clock signals generated by a temperature-compensated quartz crystal oscillator (TCXO) using ZCA-DRS.

\section{Background}\label{Sec:Review}
\begin{figure}[ht]
\includegraphics{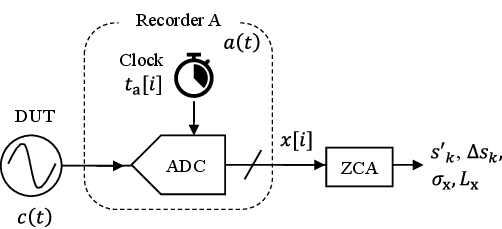}
\caption{\label{Fig:diagram_SRS} Schematic of ZCA-SRS. The DUT signal $c(t)$ is fed into recorder A to obtain the recorded waveform $x[i]$. ZCA outputs $s'_k$ and $\varDelta s_k$. }
\end{figure}

Fig. \ref{Fig:diagram_SRS} shows a schematic of zero-crossing analysis with a single recorder setup (ZCA-SRS). 
The output signal of the DUT $c(t)$ is fed into Recorder A. The sampling timing of the ADC is expressed as $t_\mathrm{a}[i]$. The additional noise caused by the recording process is represented by $a(t)$. Then, the recorded waveform $x[i]$ is given by
\begin{align}
x[i] &=c(t_\mathrm{a}[i])+a(t_\mathrm{a}[i]). \label{Eq:x}
\end{align}
The objective of zero-crossing analysis (ZCA) is to output the jitter-rejected zero-crossing timings (JRZCTs) and zero-crossing fluctuations (ZCFs). The ZCA process has four steps: (i) The continuous function $x'(t)$ is obtained by multiplying the window function $\{w[i]\}$ and $\{x[i]\}$, applying a bandpass filter using the fast Fourier transform (FFT) method, interpolating data points using the same FFT method, and connecting the interpolated points with straight lines. (ii) Zero-crossing timings $(s_1, s_2, \cdots, s_M)$ are identified by solving $x'(s_k)=0$. $M$ is the number of the zero-crossing points to derive $\sigma_{n2}$ and $L_\mathrm{c}(f)$. (iii) JRZCTs $(s'_1, s'_2, \cdots, s_M)$ are determined by using the least-squares method on the $s_k$ vs. $k$ plot. $s'_k$ is related to $k$, $\omega$, and $\theta_0$ as follows: 
\begin{align}
s'_k=\frac{(2k-1) (\pi/2)-\theta_0}{\omega}.
\end{align}
(iv) ZCFs ($\varDelta s_k$) are obtained by $\varDelta s_k:=s_k-s'_k$. The relationship between the phase noise ($\varphi_\mathrm{c}(t)$) and the ZCFs are expressed as
\begin{align}
\varDelta s_k=\frac{(-1)^k}{\omega}\left\{\varphi_\mathrm{c}(s'_k)+\varphi_\mathrm{a}(s'_k)\right\}, \label{Eq:delta_s_k}
\end{align}
where $\varphi_\mathrm{a}(t)$ is the phase noise caused by recorder A. The ZCFs are randomly distributed around zero, and they are caused by the phase noise of the DUT and recorder A. The magnitude of the distribution is numerically expressed by using the phase jitter $\sigma_\mathrm{x}$ and SSB spectrum $L_\mathrm{x}(f)$, which are obtained from the ZCFs:
\begin{align}
\sigma_\mathrm{x}&=\mathrm{dev}\left\{\varDelta s_k\right\}, \label{Eq:sigma_x}\\
L_\mathrm{x}(f)&=\frac{\omega^2}{2}\left|\mathcal{F}\left\{\varDelta s_k\right\}\right|^2, \label{Eq:L_x}
\end{align}
where $\mathrm{dev}\{~\}$ and $\mathcal{F}\{~~\}$ are the standard deviations of the data and Fourier transform, respectively. $\sigma_\mathrm{x}$ and $L_\mathrm{x}(f)$ asymptotically approach $\sigma_\mathrm{c}$ and $L_\mathrm{c}(f)$, respectively, when the noise of recorder A is sufficiently small.

\begin{figure*}[hbtp]
\includegraphics{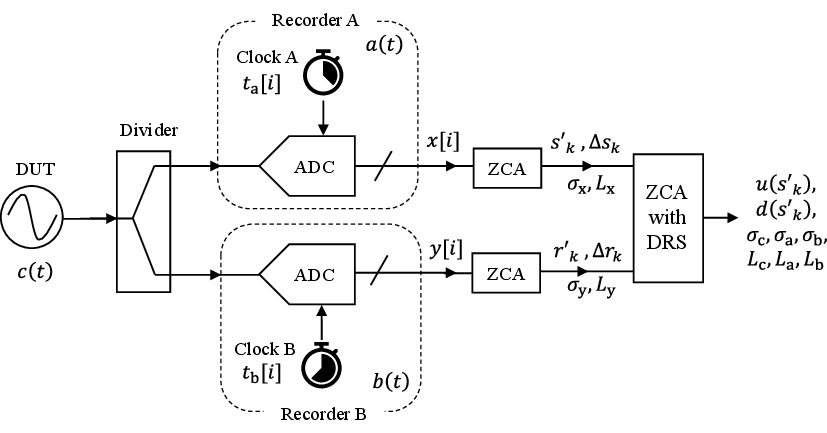}
\caption{\label{Fig:diagram_DRS} Schematic of ZCA-DRS. The DUT signal $c(t)$ is divided and fed to recorders A and B. The recorded waveforms are represented by $x[i]$ and $y[i]$. The two ZCAs output $s'_k$, $\varDelta s_k$, and $r'_k$, $\varDelta r_k$. The sum and difference of the ZCFs ($\varDelta s_k\pm\varDelta r_k$) are calculated. Finally. the phase jitter $\sigma_\mathrm{c}$ and SSB spectrum $L_\mathrm{c}(f)$ are obtained. }
\end{figure*}

Fig. \ref{Fig:diagram_DRS} shows a schematic of ZCA-DRS. 
The DUT outputs the signal $c(t)$, which is separated by a divider. The divided signals are fed to recorders A and B with the sampling timings $t_\mathrm{a}[i]$ and $t_\mathrm{b}[i]$, respectively. The internal clocks A and B of the recorders are mutually independent. The noises of recorders A and B are represented as $a(t)$ and $b(t)$, respectively. The waveforms obtained by recorders A and B are expressed as $x[i]$ and $y[i]$, respectively, where 
\begin{align}
y[i] &=c(t_\mathrm{b}[i])+b(t_\mathrm{b}[i]).
\end{align}
ZCA is applied to $x[i]$ and $y[i]$, to obtain the JRZCTs ($s'_k$ and $r'_k$) and ZCFs ($\varDelta s_k$ and $\varDelta r_k$). 
The relationship between $\varphi_\mathrm{c}(t)$ and $\varDelta s_k$ are expressed as
\begin{align}
\varDelta r_k=\frac{(-1)^k}{\omega}\left\{\varphi_\mathrm{c}(r'_k)+\varphi_\mathrm{b}(r'_k)\right\}, \label{Eq:delta_r_k_r'_k}
\end{align}
where $\varphi_\mathrm{b}(t)$ is the phase noise caused by recorder B. Because the same signals are fed into recorders A and B, the relation $s'_k=r'_k$ is established as long as the assignment of the zero-crossing index $k$ is correct. Consequently, $r'_k$ in Eq. (\ref{Eq:delta_r_k_r'_k}) can be replaced by $s'_k$ to obtain 
\begin{align}
\varDelta r_k=\frac{(-1)^k}{\omega}\left\{\varphi_\mathrm{c}(s'_k)+\varphi_\mathrm{b}(s'_k)\right\}. \label{Eq:delta_r_k}
\end{align}
The sum and difference between the two ZCFs $\varDelta s_k$ and $\varDelta r_k$ are given as follows: 
\begin{align}
u(s'_k):&=\varDelta s_k+\varDelta r_k,\label{Eq:u_def}\\
d(s'_k):&=\varDelta s_k-\varDelta r_k.\label{Eq:d_def}
\end{align}
Based on Eqs. (\ref{Eq:delta_s_k}) and (\ref{Eq:delta_r_k}), they can then be written as
\begin{align}
u(s'_k)&=\frac{(-1)^k}{\omega}\left\{2\varphi_\mathrm{c}(s'_k)+\varphi_\mathrm{a}(s'_k)+\varphi_\mathrm{b}(s'_k)\right\}, \label{Eq:u} \\
d(s'_k)&=\frac{(-1)^k}{\omega}\left\{\varphi_\mathrm{a}(s'_k)-\varphi_\mathrm{b}(s'_k)\right\}. \label{Eq:d}
\end{align}
Eq. (\ref{Eq:u}) cotains the term $2\varphi_\mathrm{c}(s'_k)$ because both $\varDelta s_k$ and $\varDelta r_k$ are shifted when the phase of the DUT is shifted. In contrast, $\varphi_\mathrm{c}(s'_k)$ vanishes and $\varphi_\mathrm{a}(s'_k)$ and $\varphi_\mathrm{b}(s'_k)$ remain in Eq. (\ref{Eq:d}). When the correlation between $\varphi_\mathrm{a}(t)$ and $\varphi_\mathrm{b}(t)$ is negligible, the phase jitter $\sigma_\mathrm{c}$ and SSB spectrum $L_\mathrm{c}(f)$ are obtained as follows: 
\begin{align}
\sigma_\mathrm{c} &=\frac{1}{2}\sqrt{\mathbb{V}\{u(s'_k)\}-\mathbb{V}\{d(s'_k)\}},\label{Eq:sigma_n2}\\
L_\mathrm{c}(f) &=\frac{\omega^2}{8}\left(\left|\mathcal{F}\{u(s'_k)\}\right|^2-\left|\mathcal{F}\{d(s'_k)\}\right|^2\right), \label{Eq:L}
\end{align}
where $\mathbb{V}\{~\}$ denotes the variance of the data.

After the phase noise of the DUT is calculated, the additional phase noise caused by the recording process (i.e., $\varphi_\mathrm{a}(t)$ and $\varphi_\mathrm{b}(t)$ in Eqs. (\ref{Eq:delta_s_k}) and (\ref{Eq:delta_r_k})) can be evaluated. Based on the addition rule of uncertainty, the phase jitter of recorder A or B ($\sigma_{a}$ or $\sigma_\mathrm{b}$, respectively) and SSB spectrum of recorder A or B ($L_\mathrm{a}(f)$ or $L_\mathrm{b}(f)$, respectively) are expressed by 
\begin{align}
\sigma_\mathrm{a,b}&=\sqrt{(\sigma_\mathrm{x,y})^2-(\sigma_\mathrm{c})^2},\\
L_\mathrm{a,b}(f)&=L_\mathrm{x,y}(f)-L_\mathrm{c}(f), \label{Eq:L_ab}
\end{align}
where $\sigma_\mathrm{y}$ and $L_\mathrm{y}(f)$ are the phase jitter and SSB spectrum obtained from $\varDelta r_k$. Similarly to Eqs. (\ref{Eq:sigma_x}) and (\ref{Eq:L_x}), these are expressed as follows:
\begin{align}
\sigma_\mathrm{y}&=\mathrm{dev}\left\{\varDelta r_k\right\}, \label{Eq:sigma_y}\\
L_\mathrm{y}(f)&=\frac{\omega^2}{2}\left|\mathcal{F}\left\{\varDelta r_k\right\}\right|^2. \label{Eq:L_y}
\end{align}
The phase noise of the recorder represents the actual performance of an instrument for a continuous-wave input. This is beneficial for selecting the most suitable oscilloscope model among those available. With ZCA-DRS, we can evaluate the performances of both the signal generator and measurement instruments simultaneously. We obtain the phase noises of the instruments ($L_\mathrm{a}$ and $L_\mathrm{b}$) from the difference between the phase noises of the waveforms recorded by the oscilloscopes ($L_\mathrm{x}$ and $L_\mathrm{y}$) and the DUT ($L_\mathrm{c}$). 

\section{Experimental}\label{Sec:Experiment}
\begin{figure}
\includegraphics{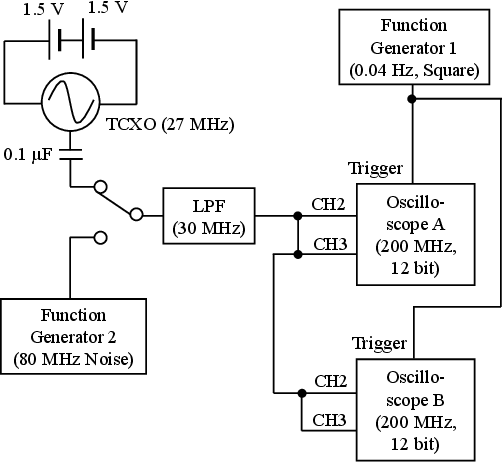}
\caption{\label{Fig:experiment} Experimental apparatus. The TCXO was driven by two dry batteries (1.5V). The DC and high-harmonic components were eliminated by a capacitor (0.1 $\mathrm{\rmu F}$) and LPF. The output signal was fed to oscilloscopes A and B. The trigger timings of the oscilloscopes were controlled by using function generator 1. Function Generator 2 was used to calibrate the experimental parameters. }
\end{figure}

Fig. \ref{Fig:experiment} shows the apparatus used in this study. 
A sinusoidal signal was generated by a TCXO (TG3225CEN, EPSON) at 27 MHz. The electric power was supplied by two dry batteries ($1.5~\mathrm{V}$). A capacitor ($0.1~\mathrm{\rmu F}$) was inserted to block the DC component of the output signal. An LPF with a cutoff frequency of $30~\mathrm{MHz}$ was inserted to block the high-harmonic components. The output of the LPF was fed to Channels 2 and 3 of oscilloscopes A and B (WaveSurfer 4024HD, Teledyne LeCroy) with a bandwidth and A/D resolution of $200~\mathrm{MHz}$ and 12 bits, respectively. The LPF and oscilloscope inputs were connected by standard Bayonet Neill-Concelman (BNC) cables and tee connectors. To absorb the transmitted power at the most distant terminal, the input impedance of channel 3 in oscilloscope B was set at $50~\mathrm{\Omega}$, while those of the other channels were set at $1~\mathrm{M\Omega}$. The trigger timings of oscilloscopes A and B were controlled by feeding square waves generated by a function generator (33509B, Keysight Technologies) into external trigger inputs. The lengths of the two BNC cables between the function generator and external trigger inputs were adjusted so that the same signals were recorded. The deskew parameters between channels 2 and 3 in oscilloscopes A and B were adjusted in the same manner. To calibrate the lengths of the cables and deskew parameters, a function generator (33612A, Keysight Technologies) that output a random noise signal was connected before the LPF instead of the TCXO. The sampling rate and measurement duration of the oscilloscopes were set at $F_\mathrm{s}=2.5~\mathrm{GSa/s}$ and $T=10~\mathrm{ms}$, respectively. The oscilloscopes repeatedly acquired the signal and stored the displayed waveforms to microSD cards once a trigger signal was received. The frequency of the trigger was set sufficiently low at $0.04~\mathrm{Hz}$ to avoid failure during each saving process. After the measurements were completed, the microSD cards were ejected, and the data were copied to a computer for ZCA. 

The average of the waveforms obtained by channels 2 and 3 in oscilloscope A was treated as $x[i]$ and the average waveform obtained by channel 2 and 3 in oscilloscope B was treated as $y[i]$. As shown in Fig. \ref{Fig:diagram_DRS}, ZCA was independently applied to $x[i]$ and $y[i]$. The window function multiplied before the FFT had the shape of an isosceles trapezoid with equal rising and falling times. The bandwidth constraint of $f_\mathrm{C}-B_\mathrm{W}\leqslant f \leqslant f_\mathrm{C}+B_\mathrm{W}$ was applied to the waveforms, where $f_\mathrm{C}:=\omega/(2\pi)=27~\mathrm{MHz}$ was the central frequency of the TCXO, and $B_\mathrm{W}=10~\mathrm{MHz}$ was the bandwidth of interest. As shown in Fig. \ref{Fig:diagram_DRS}, the JRZCTs and ZCFs obtained by oscilloscope A were represented by $s'_k$ and $\varDelta s_k$, and those obtained by oscilloscope B were represented by $r'_k$ and $\varDelta r_k$, respectively. The sum and difference of the ZCFs, (i.e., $u(s'_k)$ and $d(s'_k)$) were calculated by using Eqs. (\ref{Eq:u_def}) and (\ref{Eq:d_def}), respectively. Finally, the phase jitter $\sigma_\mathrm{c,a,b}$ and SSB spectrum $L_\mathrm{c,a,b}(f)$ were obtained by using Eqs. (\ref{Eq:sigma_x}), (\ref{Eq:L_x}) and (\ref{Eq:sigma_n2})--(\ref{Eq:L_y}). Note that the number of JRZCT and ZCF obtained for each measurement was approximately $M\approx 4.8\times 10^5$. 

\section{Results}\label{Sec:Result}
\subsection{Phase Jitters}
\begin{figure}[htbp]
\includegraphics{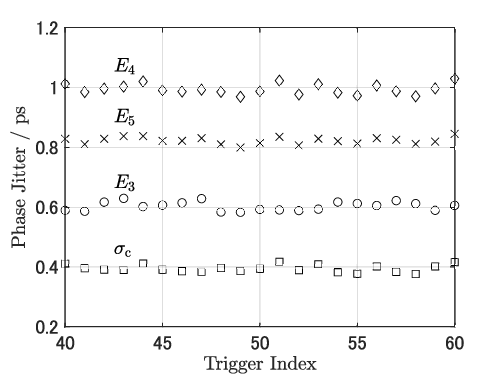}
\caption{\label{Fig:TDA_CMOS3} Phase jitters of the double recorders ($E_3$, circle), recorded waveforms ($E_4$, diamond), and DUT ($\sigma_\mathrm{c}$, square) for each measurement. For comparison, the phase jitter when the ZCA with DRS does not work correctly ($E_5$, cross) is also plotted. }
\end{figure}

The measured phase jitters of different triggers wer obtaind from the ZCFs of the recorderd waveforms ($\varDelta s_k$ and $\varDelta r_k$). The parameter $E_3$ corresponds to the phase jitter generated by the double recorders. 
\begin{align}
E_3&:=\mathrm{dev}\{\varDelta s_k-\varDelta r_k\}.
\end{align}
The parameter $E_4$ corresponds to the phase jitter observed at the double recorders, which contains the phase jitter of the DUT and double recorders: 
\begin{align}
E_4&:=\mathrm{dev}\{\varDelta s_k+\varDelta r_k\}.
\end{align}
The parameter $E_5$ corresponds to the random average of $\varDelta s_k$ and $\varDelta r_k$: 
\begin{align}
E_5&:=\sqrt{\mathbb{V}\{\varDelta s_k\}+\mathbb{V}\{\varDelta r_k\}}.
\end{align}
These parameters satisfy $E_3<E_5<E_4$ when ZCA-DRS works correctly. The phase jitter of the DUT $\sigma_\mathrm{c}$ can be calculated from the difference between $E_3$ and $E_4$. 
\begin{align}
\sigma_\mathrm{c}&=\frac{1}{2}\sqrt{(E_4)^2-(E_3)^2}.
\end{align}
Fig. \ref{Fig:TDA_CMOS3} shows the measured phase jitters of each trigger, where the horizontal axis represents the trigger index. $E_3$, $E_4$, $E_5$, and $\sigma_\mathrm{c}$ are plotted as cicles, diamonds, crosses, and squares, respectively.

The relationship $E_3<E_4$ was observed for each measurement, which indicates that ZCA-DRS was successful because the DUT contributes to $E_4$ and not to $E_3$. The relationship $E_3<E_5<E_4$ was observed, which was also expected. Finally, the relationship $\sigma_\mathrm{c}<E_3$ was ovserved, which indicates that the DUT had less phase noise than the measurement instruments. Hence, ZCA-DRS was essential because it allows the noise of DUT to be measured even when it is less than that of the measurement instruments. The phase jitter of the DUT between each measurement fluctuated slightly. The mean value and standard uncertainty for $m=100$ measurements were 
\begin{align}
\langle \sigma_\mathrm{c}\rangle_m=0.395(16)~\mathrm{ps},
\end{align}
where $\langle~~\rangle_m$ denotes the mean of $m$ values and the number in parentheses represents the standard deviation of the mean. The offset frequency range of the phase jitter was limited to $1/T \leqslant f\leqslant B_\mathrm{W}$, or $100~\mathrm{Hz}\leqslant f\leqslant 10~\mathrm{MHz}$. 

\subsection{Phase Noise}
\begin{figure}[htbp]
\includegraphics{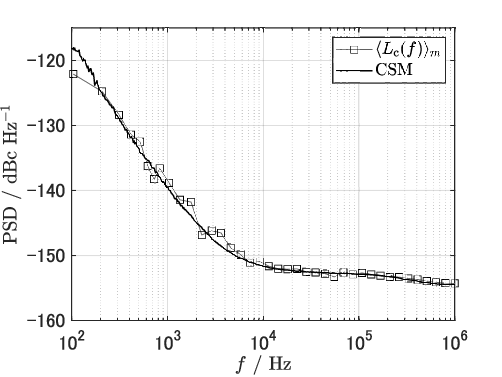}
\caption{\label{Fig:Spectrum_CMOS3} SSB spectrum measurements obtained by ZCA-DRS (squares) and CSM (solid line). }
\end{figure}

Fig. \ref{Fig:Spectrum_CMOS3} shows the SSB spectrum measurements.
The horizontal axis represents the offset frequency. The power spectral density (PSD) is plotted by square markers connected by gray lines. The PSD was obtained by averaging over $m=100$ measurements, after which the data points were resampled for ease of view in a log-log scale plot. For comparison, the PSD obtained by CSM using commercial instruments (E5052B, Keysight) is drawn as a solid line. The ZCA-DRS results agreed with the CSM result at $f\geqslant 200~\mathrm{Hz}$, which indicates that ZCA-DRS was successfully performed. The difference from the CSM results at $f< 200~\mathrm{Hz}$ can be attributed to the frequency resolution of the ZCA-DRS. Based on Eq. (\ref{Eq:L}), the frequency resolution is given by 
\begin{align}
\varDelta f=\frac{2f_\mathrm{C}}{M}.
\end{align}
The actual value was $\varDelta f\approx 104~\mathrm{Hz}$.

\subsection{Phase noise components of DUT and a recorder}\label{SubSec:Spectrum_Oscillo}
\begin{figure}[htbp]
\includegraphics{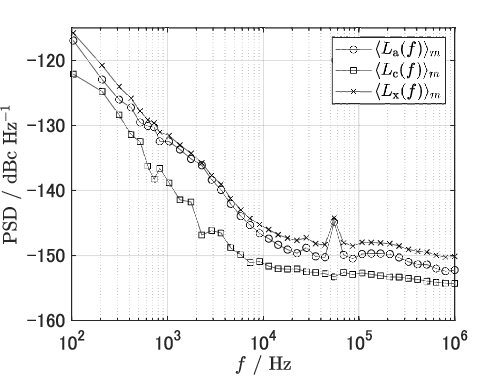}\\
\caption{\label{Fig:Spectrum_Oscillo} SSB phase noise components obtained by recorder A. The total noise ($\langle L_\mathrm{x}\rangle_m$), noise of the DUT ($\langle L_\mathrm{c}\rangle_m$) and noise of recorder A ($\langle L_\mathrm{a}\rangle_m$) are plotted. The structure exhibited broad excess noise at $1~\mathrm{kHz} \leqslant f \leqslant 6~\mathrm{kHz}$ and a sharp peak at $f=55~\mathrm{kHz}$. }
\end{figure}

Fig. \ref{Fig:Spectrum_Oscillo} shows the components of the phase noise obtained by recorder A. 
The phase noise of the DUT $\langle L_\mathrm{c}\rangle_m$ is also shown for comparison. The SSB spectrum of the recorded waveform obtained by recorder A ($\langle L_\mathrm{x}\rangle_m$) is plotted by crosses. The SSB spectra of the DUT ($\langle L_\mathrm{c}\rangle_m$) and recorder A ($\langle L_\mathrm{a}\rangle_m$) are plotted by squares and circles, respectively. The phase noise of recorder A $L_\mathrm{a}$ was obtained by $L_\mathrm{x}-L_\mathrm{c}$ as shown in Eq. (\ref{Eq:L_ab}). In all bands, the DUT had less phase noise than recorder A (i.e., $\langle L_\mathrm{c}\rangle_m<\langle L_\mathrm{a}\rangle_m$). This shows that the phase noise of the DUT cannot be obtained by a single recorder (i.e., ZCA-SRS) and that two recorders are essential (i.e., ZCA-DRS). The recorder noise was prominent in the region of $1~\mathrm{kHz} \leqslant f \leqslant 6~\mathrm{kHz}$, where $\langle L_\mathrm{a}\rangle_m$ did not decrease according to the power law and the difference between $\langle L_\mathrm{c}\rangle_m$ and $\langle L_\mathrm{a}\rangle_m$ exceeded $6~\mathrm{dBc/Hz}$. In addition, a sharp peak was observed at $f=55~\mathrm{kHz}$. This peak was successfully removed by using ZCA-DRS. 

\subsection{Measurement Uncertainty}
\begin{figure}[htbp]
\includegraphics{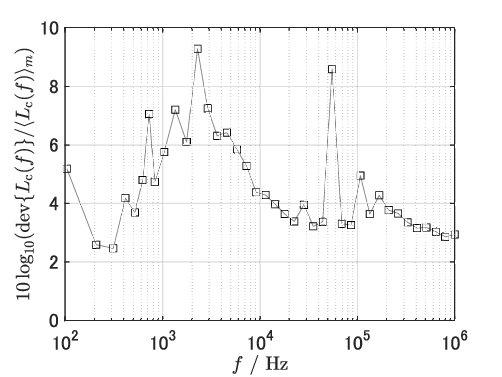}
\caption{\label{Fig:Uncertainty_CMOS3} Standard deviation of the PSD values $\mathrm{dev}\{L_\mathrm{c}(f)\}$ vs. offset frequency $f$. The standard deviation was normalized by  the mean value $\langle L_\mathrm{c}(f)\rangle_m$ and expressed in decibels. The normalized standard deviation $\mathrm{dev}\{L_\mathrm{c}(f)\}/\langle L_\mathrm{c}(f)\rangle_m$ was distributed from $10^{2.5/10}$ to $10^{9.3/10}$. }
\end{figure}

We estimated the uncertainty of ZCA-DRS when applied to measuring the phase noise of a TCXO. the measurement. Figure \ref{Fig:Uncertainty_CMOS3} shows the standard deviation of the PSD values for each trigger normalized by the mean, which we call the relative uncertainty. 
When the measurement was repeated $m=100$ times, the relative uncertainty is reduced to less than 1. For each offset frequency, the relative uncertainty was distributed between 2.5 dB and 9.3 dB.  The measurement uncertainty of the measurement was 10 dB below the relative uncertainty because the measurement was repeated for $m=100$ times. Therefore, the uncertainty of the measurements shown in Fig. \ref{Fig:Spectrum_CMOS3} was less than 0 dB. This is the uncertainty of the 1 s measurement. Thus, this can be improved by increasing the measurement time.

\section{Discussions}\label{Sec:Discussion}

\subsection{Origin of Noise in Measurement Instruments}\label{SubSec:noise_oscillo}
\begin{figure}[htbp]
\includegraphics{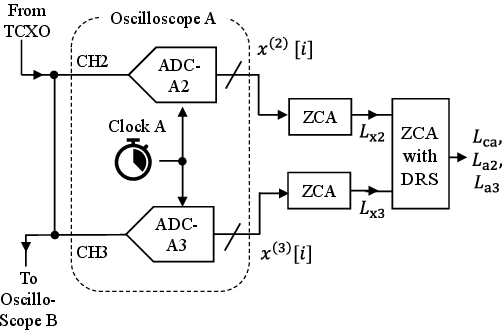}
\caption{\label{Fig:diagram_discuss} Internal structure of the oscilloscope and procedure used to analyze the origin of phase noise. }
\end{figure}

\begin{figure}[htbp]
\includegraphics{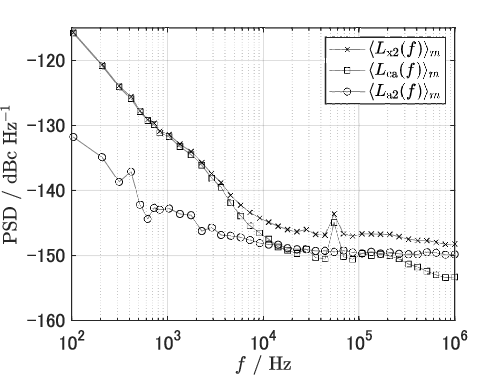}
\caption{\label{Fig:Spectrum_CMOS3_ClockA} Phase noise originating from the TCXO and internal clock embedded in oscilloscope A ($L_\mathrm{ca}$) and from the ADC in oscilloscope A ($L_\mathrm{a2}$). The phase noise of the waveform obtained from channel 2 of oscilloscope A ($L_\mathrm{x2}$) is also plotted, which is approximately 3 dB higher than $\langle L_\mathrm{x}\rangle_m$ in Fig. \ref{Fig:Spectrum_Oscillo}. }
\end{figure}

We analyze the origin of the phase noise in oscilloscope as follows. As shown in Fig. \ref{Fig:diagram_discuss}, identical electrical signals were fed into channels 2 and 3 of oscilloscope A. 
The phase noise obtained using $x^{(2)}[i]$ and $x^{(3)}[i]$ are denoted by $L_\mathrm{x2}$ and $L_\mathrm{x3}$, respectively. The phase noise of TCXO and the internal clock of oscilloscope A is denoted by $L_\mathrm{ca}$. The phase-independent (PI) noise caused by the ADC of channels 2 and 3 is denoted by $L_\mathrm{a2}$ and $L_\mathrm{a3}$, respectively. These parameters and $L_\mathrm{x}$ have the following relationships: 
\begin{align}
L_\mathrm{x}(f)&=L_\mathrm{ca}(f)+\frac{L_\mathrm{a2}(f)+L_\mathrm{a3}(f)}{4}\\
&=\frac{L_\mathrm{ca}(f)}{2}+\frac{L_\mathrm{x2}(f)+L_\mathrm{x3}(f)}{4},\\
L_\mathrm{x2}(f)&=L_\mathrm{ca}(f)+L_\mathrm{a2}(f),\\
L_\mathrm{x3}(f)&=L_\mathrm{ca}(f)+L_\mathrm{a3}(f).
\end{align}
Fig. \ref{Fig:Spectrum_CMOS3_ClockA} shows the analysis results. 
The primary component of $L_\mathrm{x2}$ was $L_\mathrm{ca}$ in the low-frequency band ($f \leqslant 20~\mathrm{kHz}$) and $L_\mathrm{a2}$ in the high-frequency band ($f>20~\mathrm{kHz}$). Thus, the main source of the phase noise was the sampling jitter when $f \leqslant 20~\mathrm{kHz}$ and the PI noise produced by the ADC when $f>20~\mathrm{kHz}$. The internal clock of the oscilloscope was the main origin of the measured phase noise, and the noise of the ADC was sufficiently small. The peak for $L_\mathrm{x2}$ at $f=55~\mathrm{kHz}$ can be attributed to the internal clock of the oscilloscope Therefore, suppressing the sampling jitter of the oscilloscope is important. 

\subsection{Detection Limit of ZCA-SRS}
In Section \ref{Sec:Review}, we noticed that recent progress in A/D converters enables us to directly observe the phase noise of a low-noise oscillator with ZCA. To clarify the requirement for the A/D converter, we derive the quantization noise floor level for ZCA-SRS, which is simply determined by the A/D resolution of the measurement instrument and the sampling rate. We assumed that the sinusoidal wave expressed in Eq. (\ref{Eq:c}) was recorded by an instrument with an A/D resolution of $N$ and sampling rate of $F_\mathrm{s}$. Then, the formula of Eq. (10) in the reference\cite{Kester09} can be used to predict the signal-to-noise ratio $R_\mathrm{sn}$
\begin{align}
\frac{R_\mathrm{sn}}{\mathrm{dB}}=\frac{6.02 N}{\mathrm{bit}}+1.76+10\log_{10}\left(\frac{F_\mathrm{s}/2}{f_\mathrm{z}}\right), \label{Eq:R_sn}
\end{align}
where $f_\mathrm{z}$ is the bandwidth of the ZCA. The third term on the right-hand side is the process gain. The signal-to-noise ratio of ZCA ($R_\mathrm{zca}$) is twice that of $R_\mathrm{sn}$:
\begin{align}
R_\mathrm{zca}=2R_\mathrm{sn}. \label{Eq:R_zca}
\end{align}
This is because ZCA is sensitive to temporally varying noise whose phase is orthogonal to the carrier wave, as expressed by Eq. (4) in our primary study.\cite{Takeuchi2023} In other words, ZCA is not sensitive to the in-phase component of the temporally varying noise, which is defined as the amplitude modulation(AM) $\alpha(t)$ in Eq. ~ (\ref{Eq:c}). The quantization noise represented as the PSD is expressed by
\begin{align}
L_\mathrm{min}=\frac{(R_\mathrm{zca})^{-1}}{f_\mathrm{z}}. \label{Eq:L_min}
\end{align}
Therefore, the following formula is obtained:
\begin{align}
\frac{L_\mathrm{min}}{\mathrm{dBc/Hz}}
=-\frac{6.02 N}{\mathrm{bit}}-1.76-10\log_{10}\left(\frac{F_\mathrm{s}}{\mathrm{Sa/s}}\right). \label{Eq:L_min_formula}
\end{align}
This formula is beneficial for estimating the detection limit using ZCA-SRS. Note that Eq. (\ref{Eq:L_min_formula}) does not include $f_\mathrm{z}$. The sampling rate was fixed to $F_\mathrm{s}=2.5~\mathrm{GSa/s}$ in this study. Based on Eq. (\ref{Eq:L_min_formula}), we can make two insights. First, the detection limit for $N=10~\mathrm{bit}$ is 
\begin{align}
L_\mathrm{min}(10~\mathrm{bit})&=-156~\mathrm{dBc/Hz}.
\end{align}
As shown in Fig. \ref{Fig:Spectrum_CMOS3}, the noise floor of the TCXO is $\mathrm{min}\{\langle L_\mathrm{c}\rangle_m\} \approx 154~\mathrm{dBc/Hz}$, where $\mathrm{min}\{~\}$ denotes the minimum value in the frequency region $f \leqslant 1~\mathrm{MHz}$. These two values are close. Hence, the effective A/D resolution of the instrument requires $N\gtrsim 10~\mathrm{bit}$ for phase noise measurement of a TCXO. It is difficult to observe phase noise by using an oscilloscope with an A/D resolution of $N_\mathrm{res}=8~\mathrm{bit}$. Thus, oscilloscopes with an A/D resolution of $N_\mathrm{res}=12~\mathrm{bit}$ are more promising. Second, the detection limit for $N=9~\mathrm{bit}$ is 
\begin{align}
L_\mathrm{min}(9~\mathrm{bit})&=-150~\mathrm{dBc/Hz}.
\end{align}
As shown in Fig. \ref{Fig:Spectrum_CMOS3_ClockA}, the noise floor of the ADC was $\mathrm{min}\{\langle L_\mathrm{a2}\rangle_m\} \approx 150~\mathrm{dBc/Hz}$. The two values are similar. This suggests that the effective A/D resolution of the oscilloscope used in this study was $N\approx 9~\mathrm{bit}$. This is less than 10 bit, which we previously showed is the needed to measure the phase noise of a TCXO. Hence, ZCA-DRS was essential. 

\section{Concluding remarks}\label{Sec:Conclusion}
We extended ZCA-DRS to the RF range by using two high-speed digital oscilloscopes and successfully measured the phase noise of a TCXO. Measurements were repeated $m=100$ times to reduce uncertainty. The obtained phase noise spectrum agreed well with the CSM results. We used ZCA-DRS to evaluate the noise performance of the measurement instruments at frequency bands close to the carrier frequency. Excessive noise in the oscilloscope was observed at offset frequencies of $1~\mathrm{kHz}\leqslant f\leqslant 6~\mathrm{kHz}$ and $f\approx 55~\mathrm{kHz}$, which we identified as originating from the sampling jitter of the oscilloscope. A simple analysis confirmed that measurements were difficult for oscilloscope with an A/D resolution of 8 bits. The A/D resolution needed to be increased to 12 bit to measure the phase noise of the TCXO. Measuring lower phase noise requires instruments with less sampling jitter and ADC noise. Future research will involve improving the detection limit. 

\section*{Acknowledgments}
We thank EPSON for providing the samples and phase-noise data for the TCXO.

\section*{Author declarations}
\subsection*{Conflicts of interest}
The authors have no conflicts to disclose. 

\subsection*{Author contributions}
\textbf{Makoto Takeuchi:}
 Conceptualization(equal); Data curation (lead); Formal analysis (lead);  Funding acquisition (supporting); Investigation (lead); Methodology (lead); Project administration (supporting); Resources (supporting); Software (supporting); Supervision (supporting); Validation (lead); Visualization (lead); Writing – original draft (lead); Writing – review \& editing (supporting).
\textbf{Haruo Saito:}
 Conceptualization(equal); Data curation (equal); Formal analysis (equal); Funding acquisition (lead); Investigation (equal); Methodology (equal); Project administration (lead); Resources (lead); Software (lead); Supervision (lead); Validation (equal); Visualization (supporting); Writing – original draft (supporting); Writing – review \& editing (lead).

\section*{Data availability}
The data that support the findings of this study are available from the corresponding authors upon reasonable request. 



%
%

%


\providecommand{\noopsort}[1]{}\providecommand{\singleletter}[1]{#1}%

\end{document}